\documentclass[trackchanges,twocolumn]{aastex701}

\usepackage{bm}
\usepackage{amsmath}
\usepackage{multirow} 
\usepackage{booktabs}

\begin{document}

\title{Hydrodynamical simulation of wind production from hot accretion flows in tidal disruption events}

\correspondingauthor{Mingjun Liu, De-Fu Bu}
\email{mjliu@bao.ac.cn, dfbu@shnu.edu.cn}

\author[0009-0009-8982-2361]{Mingjun Liu}
\affiliation{National Astronomical Observatories, Chinese Academy of Sciences, 20A Datun Road, Beijing 100101, China}
\email{mjliu@bao.ac.cn}

\author[0000-0002-0427-520X]{De-Fu Bu} 
\affiliation{Shanghai Key Lab for Astrophysics, Shanghai Normal University, 100 Guilin Road, Shanghai 200234, China}
\email{dfbu@shnu.edu.cn}

\author[0000-0002-2419-9590]{Xiao-Hong Yang}
\affiliation{Department of Physics, Chongqing University, Chongqing 400044, China}
\email{yangxh@cqu.edu.cn}

\author{Jiaqi Li}
\affiliation{School of Physics and Astronomy, Beijing Normal University, Beijing 100875, China}
\email{202431101075@mail.bnu.edu.cn}

\author[0000-0003-4200-9954]{Huaqing Cheng}
\affiliation{National Astronomical Observatories, Chinese Academy of Sciences, 20A Datun Road, Beijing 100101, China}
\email{hqcheng@nao.cas.cn}

\author{Qinyu Wu}
\affiliation{National Astronomical Observatories, Chinese Academy of Sciences, 20A Datun Road, Beijing 100101, China}
\affiliation{School of Astronomy and Space Science, University of Chinese Academy of Sciences, 19A Yuquan Road, Beijing 100049, China}
\email{qywu@bao.ac.cn}

\author[0009-0003-9214-7316]{Wenjie Zhang}
\affiliation{National Astronomical Observatories, Chinese Academy of Sciences, 20A Datun Road, Beijing 100101, China}
\email{zhangwj@bao.ac.cn}

\author[0000-0001-9920-4019]{B. F. Liu}
\affiliation{National Astronomical Observatories, Chinese Academy of Sciences, 20A Datun Road, Beijing 100101, China}
\affiliation{School of Astronomy and Space Science, University of Chinese Academy of Sciences, 19A Yuquan Road, Beijing 100049, China}
\email{bfliu@nao.cas.cn}

\begin{abstract}

Wind is a key mechanism for supermassive black hole (SMBH) feedback to their host galaxies. In tidal disruption events (TDEs), black holes spend most of their time accreting at highly sub-Eddington rates, implying that feedback from persistent sub-Eddington winds could be significant. We investigate the effects of black hole mass, viscosity parameter and stellar debris temperature on the properties of winds from hot accretion flows in TDEs. We find that more massive black holes yield a higher accreted fraction and launch faster winds, while the debris temperature has a negligible influence on the accretion flow. For $\alpha=0.1$, the mildly-relativistic unbound winds ($\sim 0.1c$) are launched predominantly from the outside of the accretion flows along the equatorial plane, with a kinetic energy of $\sim10^{-4}L_\mathrm{Edd}$. In contrast, convective bound outflows dominate for $\alpha=0.01$, which differs from the true winds typically seen in active galactic nuclei and X-ray binaries. Potential applications for explaining delayed radio brightening in TDEs at $\sim10^3$ days and for searching for intermediate-mass black holes through radio and X-ray surveys are also discussed.

\end{abstract}

\keywords{\uat{Accretion}{14} --- \uat{Astrophysical fluid dynamics}{101} --- \uat{Black holes}{162} --- \uat{High Energy astrophysics}{739} --- \uat{Tidal disruption}{1696}}


\section{Introduction} \label{sec:intro}

The co-evolution of supermassive black holes (SMBHs) and their host galaxies remains one of the central challenges in the study of galaxy formation. Disk winds play a critical role in linking SMBH activity to galactic processes, such as active galactic nuclei (AGN) feedback \citep[see][and references therein]{Kormendy2013,King2015}. These winds ubiquitously exist from super to sub-Eddington accretion regimes \citep[e.g.,][]{Ayal2000,Strubbe2009,Yuan2014,Bu2025}, including slim disk \citep[e.g.,][]{Abramowicz1988}, standard thin disk \citep[e.g.,][]{Shakura1973,Novikov1973} and hot accretion flows \citep[e.g.,][]{Narayan1994,Narayan1995a,Narayan1995b}.

Among various SMBH activities, tidal disruption events \citep[TDEs, e.g.,][]{Hills1975,Nolthenius1982,Rees1988} offer a unique probe for accretion physics, as the fallback rate of stellar debris declines from super-Eddington to sub-Eddington over months to years. Winds launched during the super-Eddington phase of TDEs have been widely studied. They are thought to be driven by processes such as radiation pressure in the slim disk \citep[e.g.,][]{Ohsuga2005,Poutanen2007,Dai2018,Curd2019,Thomsen2022b} or the self‑intersection shock during disk formation \citep[e.g.,][]{Jiang2016,Lu2020}. The magnetic field is also supposed to launch super-Eddington winds \citep[e.g.,][]{Sadowski2015a,Sadowski2015b,Utsumi2022,Yang2023}. These powerful winds can produce blueshifted and broadened emission lines \citep[e.g.,][]{Strubbe2011,Thomsen2022a,Zhang2024}, contribute to the observed dichotomy between X‑ray and optical TDE populations through inclination effects \citep{Dai2018} and interact with the circumnuclear medium \citep[e.g.,][]{Matsumoto2021} and dense clouds in the torus \citep[e.g.,][]{Mou2021a} to produce delayed radio \citep[e.g.,][]{Alexander2020,Bu2023a,Zhuang2025}, X-ray \citep{Mou2021a}, Gamma-ray \citep{Mou2021b} and neutrino emissions \citep[e.g.,][]{Stein2021,Plotko2024}. Many of these features have been found, e.g., the blue-shifted reflected lines in ASASSN-14li \citep{Kara2018}, AT2022lri \citep{Yao2024} and EP240222a \citep{Jin2025}, the blue-shifted absorption lines in ASASSN-14li \citep{Cenko2016}, iPTF15af \citep{Blagorodnova2019}, iPTF16fnl \citep{Brown2018} and AT2018zr \citep{Hung2019}, as well as radio emission in dozens of TDEs \citep[e.g.,][]{Alexander2020,Horesh2021,Cendes2022,Cendes2024,Goodwin2022,Perlman2022,Sfaradi2022,Wang2023,Wang2025,Anumarlapudi2024,Zhang2024b,Lin2025}. 

\begin{figure}[ht!]
\includegraphics[width=\columnwidth]{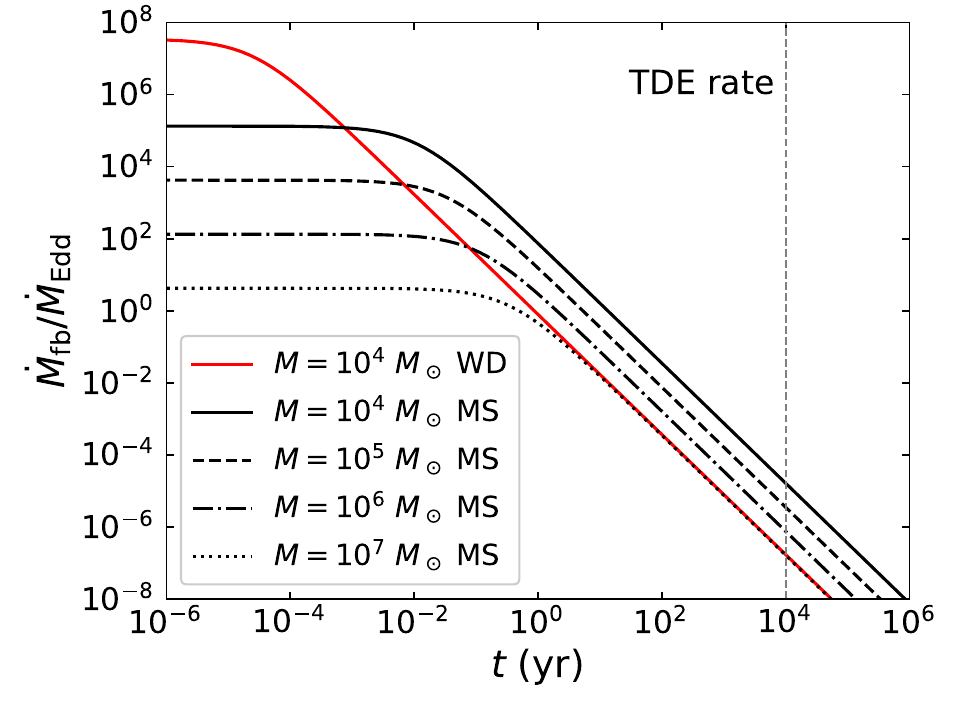}
\caption{Evolution of fallback rates for solar-type main sequence star (MS, black) and 0.6 $M_\odot$ white dwarf (WD, red) disrupted by black holes of different masses. The results for $10^4\ M_\odot$, $10^5\ M_\odot$, $10^6\ M_\odot$ and $10^7\ M_\odot$ black holes are marked by solid, dashed, dased-dotted and dotted lines, respectively. The recurrence time of TDEs corresponding to the averaged per-galaxy TDE rate \citep[$\gtrsim1\times10^{-4}$ yr$^{-1}$,][]{Stone2020} is marked by the grey dashed line.  
\label{fig:fb}}
\end{figure}

Super-Eddington winds mark the onset of black hole feedback in TDEs. As the fallback rate declines, the accretion flow transitions to a standard thin disk and eventually settles into the hot accretion flow regime. During this phase, as illustrated in Figure \ref{fig:fb}, the fallback rate requires decades to thousands of years to decline by an order of magnitude, a timescale comparable to the average per-galaxy TDE rate. Consequently, sub-Eddington winds are expected to drive persistent feedback into the galactic environment. Line-driven winds from the thin disk in TDEs have been investigated by \citet{Bu2026}, while the properties of winds launched from hot accretion flows in TDEs remain underexplored. 

Winds from hot accretion flow have been extensively studied \citep[see][and references therein]{Yuan2014}. When \citet{Narayan1994,Narayan1995a} emphasized the importance of the advection-dominated accretion flow (ADAF), they found that the positive Bernoulli parameter ($Be$) in ADAF indicates the presence of strong winds from this hot accretion flow. Then several simulations revealed the power-law radial profiles of the mass accretion rate \citep[e.g.,][]{Stone1999,Hawley2001,Machida2001,Stone2001,Hawley2002,Igumenshchev2003,Pen2003,Yuan2010,Pang2011,Yuan2012b,Yuan2012a}. Two competing models have been proposed to explain the inward decrease in inflow rates. In the adiabatic inflow–outflow solution scenario \citep[ADIOS, e.g.,][]{Blandford1999,Blandford2004,Begelman2012}, the unbound winds induce a genuine mass loss, while the gas in the convection-dominated accretion flow scenario \citep[CDAF, e.g.,][]{Narayan2000,Quataert2000} just moves convectively rather than escapes. Current studies favor ADIOS in the immediate vicinity of black holes \citep{Yuan2012b}, whereas CDAF appears to be more relevant for circumnuclear accretion in AGN environments \citep{Bu2016}. In recent years, winds from low-luminosity AGNs have been observed in several sources, e.g., Sgr A$^*$ \citep{Wang2013,Ma2019}, M81 \citep{Shi2021,Shi2024}, NGC 7213 \citep{Shi2022,Shi2024}, M87 \citep{Kino2022} and M32 \citep{Peng2020}. These winds are supposed to form X/Gamma-ray bubbles observed in both cosmological simulations \citep[e.g.,][]{Pillepich2021} and our galaxy \citep[e.g.,][]{Sarkar2024}.

However, TDEs feature a distinct gas supply mechanism. In typical hot accretion flows of AGNs, winds originate primarily in outer regions, whereas in TDEs, stellar debris is injected directly into the immediate vicinity of the black hole to form a transient, compact accretion flow. This fundamental difference suggests specific mass and angular momentum transfer processes in hot accretion flows of TDEs.

In this work, we performed hydrodynamic simulations to investigate the formation and properties of winds from the sub-Eddington hot accretion flows in TDEs. The structure of the paper is as follows. In Section \ref{sec:sim}, we describe our numerical setup and physical models. In Section \ref{sec:res}, we introduce the effects of debris temperature, black hole mass and viscosity parameter on winds. The potential observation applications are discussed in Section \ref{sec:obs}. The concolusions are summarized in \ref{sec:con}.

\section{Numerical method} \label{sec:sim}

\subsection{Equations}

Two-dimensional axisymmetric hydrodynamic simulations are carried out in spherical coordinates $(r,\theta,\phi)$ using the ZEUS-MP code \citep{Hayes2006}, following the approach of \citet{Bu2019}. The evolution of hot accretion flows in TDEs is described by the conservation laws of mass, momentum, and internal energy,
\begin{align}
    &\frac{\partial \rho}{\partial t}+\nabla\cdot\left(\rho\bm{v}\right)=0,\\
    &\frac{\partial \rho\bm{v}}{\partial t}+\nabla\cdot\left(\rho\bm{v}\bm{v}\right)=-\nabla p-\rho\nabla\Phi-\nabla\cdot\bm{T},\\
    &\frac{\partial e}{\partial t}+\nabla\cdot\left(e\bm{v}\right)=-p\nabla\cdot\bm{v}+\bm{T}^2/\mu,\label{eq:en}
\end{align}
where $\rho$ is the density, $\bm{v}$ is the velocity and $e$ is the internal energy density. The pressure is given by the ideal gas equation of state, $p=(\gamma-1)e$, with an adiabatic index $\gamma=5/3$. The gravitational field is described by the pseudo-Newtonian potential $\Phi=-GM/(r-R_\mathrm{S})$, where $G$ is the gravitational constant, $M$ is the black hole mass and $R_\mathrm{S}=2GM/c^2$ is the Schwarzschild radius, with the speed of light $c$. Radiative cooling is neglected in the energy equation (Eq. \ref{eq:en}) since we treat hot accretion flows. $\bm{T}$ is the viscous stress tensor and the dynamic viscosity $\mu=\rho\nu$ with the kinematic viscosity described through the $\alpha$-prescription $\nu=\alpha\sqrt{GMr}$. Following \citet{Stone1999}, we only consider azimuthal components in the viscous stress tensor,  
\begin{align}
    &T_{r,\phi}=\mu r\frac{\partial}{\partial r}\left(\frac{v_\phi}{r}\right),\\
    &T_{\theta,\phi}=\frac{\mu\sin{\theta}}{r}\frac{\partial}{\partial\theta}\left(\frac{v_\phi}{\sin{\theta}}\right).
\end{align}

\subsection{Initial conditions and numerical settings}

We assume a standard TDE scenario that involves the disruption of a solar-type star with stellar radius $R_*=R_\odot$ and mass $M_*=M_\odot$ by a supermassive black hole at the pericenter distance $R_\mathrm{p}$ of a parabolic orbit, i.e., the penetration factor $\beta=R_\mathrm{t}/R_\mathrm{p}=1$ where the tidal radius $R_\mathrm{t}=R_*\left(M/M_*\right)^{1/3}$. We focus on the late-time phase of TDEs, when the accretion flow transitions from a standard thin disk to a hot accretion flow, e.g., about 10-30 years for a solar-type star disrupted by a $10^6-10^7M_\odot$ black hole. At this stage, the accretion flow should have already circularized. We also assume that stellar debris can be very quickly circularized. Consequently, the conservation of angular momentum requires that a circularized accretion flow forms at the circularization radius $R_\mathrm{c}=2R_\mathrm{p}$. Therefore, we assume that the circularized stellar debris is injected into the computational domain near the circularization radius $R_\mathrm{c}$, following the fallback rate $\dot{M}_\mathrm{fb}$. The fallback timescale is given by $t_\mathrm{bf}=2^{-1/2}\pi\left(GM_*/R_*^3\right)^{-1/2}\left(M/M_*\right)^{1/2}$, and the fallback rate by $\dot{M}_\mathrm{fb}=\left(M_*/3t_\mathrm{fb}\right)\left(1+t/t_\mathrm{fb}\right)^{-5/3}$. For hot accretion flows in TDEs, the initial fallback rates in each case are set at 0.01 $\dot{M}_\mathrm{Edd}$, which represents the transition from a standard disk to a hot accretion flow at the critical accretion rate \citep[e.g.,][]{Narayan1995b,Abramowicz1995,Esin1996,Yuan2001,Li2023,Liu2025a,Liu2025b}. The fallback rate is therefore described as $\dot{M}_\mathrm{fb}=\left(M_*/3t_\mathrm{fb}\right)\left[1+\left(t_\mathrm{ini}+t\right)/t_\mathrm{fb}\right]^{-5/3}$, where the initial physical time $t_\mathrm{ini}$ satisfies $0.01\dot{M}_\mathrm{Edd}=\left(M_*/3t_\mathrm{fb}\right)\left(1+t_\mathrm{ini}/t_\mathrm{fb}\right)^{-5/3}$.

Since it takes only 10 days for the debris to adiabatically cool to the hydrogen recombination temperature \citep{Coughlin2016}, we assume that the debris is injected with a temperature of $T_\mathrm{inj}=1\times10^4$ K, representing an adiabatic evolution of stellar debris. For comparison, we conducted simulations with $T_\mathrm{inj}=2\times10^7$ K, the temperature that ensures hydrostatic equilibrium inside the original star \citep{Bonnerot2021}, to represent the isothermal evolution of the debris.  The parameters for each simulation are summarized in Table \ref{tab:para}.

The inner and outer radial boundaries of these simulations are set as outflow boundaries at 1.2 $R_\mathrm{S}$ and 1000 $R_\mathrm{S}$, respectively. An axisymmetric boundary condition is applied along the rotational axis. To ensure numerical convergence, the computational domain is initialized with a tenuous background gas, with density $\rho_\mathrm{ini}=10^{-23}$ g cm$^{-3}$, pressure $p_\mathrm{ini}=GM\rho/r\left(\gamma-1\right)$ and velocity $\bm{v}_\mathrm{ini}=(0,0,10^{-10}\sqrt{GM/r})$. The properties of this background gas are deliberately distinct from those of the stellar debris to conveniently verify the correct implementation of the debris injection. Although this background gas is not in hydrodynamical equilibrium, its extremely low density ensures that it has no influence on our results. Following \citet{Bu2022}, gas is injected uniformly into the region $R_\mathrm{c}-2R_\mathrm{S}\leq r\leq R_\mathrm{c}+2R_\mathrm{S}$ and $9\pi/20\leq\theta\leq11\pi/20$ surrounding the circularization radius $R_\mathrm{c}=2R_\mathrm{t}$, with Keplerian rotation $\bm{v}_\mathrm{inj}=(0,0,\sqrt{GM/r})$. The grid resolution is set to $N_r\times N_\theta=192\times88$.

\begin{table*}
    \caption{Setup of Simulation Parameters}
    \label{tab:para}
    \begin{tabular*}{\linewidth}{@{}@{\extracolsep{\fill}}cccccccc@{}}
    \hline
    \rule[-1ex]{0pt}{4ex} Models & $M_\mathrm{BH}/M_\odot$ & $\alpha$ & $T_\mathrm{inj}$ (K) & $R_\mathrm{c}/R_\mathrm{S}$ & $t_\mathrm{ini}$ (yr) & $t_\mathrm{acc}$ (day) & Be/$c^2$ at $R_\mathrm{c}$ \\
    \rule[-1ex]{0pt}{4ex} (1) & (2) & (3) & (4) & (5) & (6) & (7) & (8)\\
    \hline
    \rule[-1ex]{0pt}{3.5ex} M6$\alpha$001T4 & $10^6$ & 0.01 & $1\times10^4$ & 47.1 & 33.4 & 10.0 & $-5.5376\times10^{-3}$ \\
    \rule[-1ex]{0pt}{3.5ex} M6$\alpha$001T7 & $10^6$ & 0.01 & $2\times10^7$ & 47.1 & 33.4 & 10.0 & $-5.5329\times10^{-3}$ \\
    \rule[-1ex]{0pt}{3.5ex} M6$\alpha$01T4 & $10^6$ & 0.1 & $1\times10^4$ & 47.1 & 33.4 & 1.0 & $-5.5376\times10^{-3}$ \\
    \rule[-1ex]{0pt}{3.5ex} M6$\alpha$01T7 & $10^6$ & 0.1 & $2\times10^7$ & 47.1 & 33.4 & 1.0 & $-5.5329\times10^{-3}$ \\
    \rule[-1ex]{0pt}{3.5ex} M7$\alpha$001T4 & $10^7$ & 0.01 & $1\times10^4$ & 10.1 & 13.0 & 10.0 & $-3.0020\times10^{-2}$ \\
    \rule[-1ex]{0pt}{3.5ex} M7$\alpha$001T7 & $10^7$ & 0.01 & $2\times10^7$ & 10.1 & 13.0 & 10.0 & $-3.0015\times10^{-2}$ \\
    \rule[-1ex]{0pt}{3.5ex} M7$\alpha$01T4 & $10^7$ & 0.1 & $1\times10^4$ & 10.1 & 13.0 & 1.0 & $-3.0020\times10^{-2}$ \\
    \rule[-1ex]{0pt}{3.5ex} M7$\alpha$01T7 & $10^7$ & 0.1 & $2\times10^7$ & 10.1 & 13.0 & 1.0 & $-3.0015\times10^{-2}$ \\
    \hline
    \end{tabular*}
    \footnotesize{(1) Model; (2) Black hole mass scaled by solar mass; (3) Viscosity parameter; (4) Temperature of stellar debris; (5) Gas injection radius scaled by $R_\mathrm{S}$; (6) Initial physical time corresponding to $\dot{M}_\mathrm{fb}=0.01\dot{M}_\mathrm{Edd}$; (7) Accretion timescale $t_\mathrm{acc}=-R_\mathrm{c}/v_r$} where the radial velocity is estimated via the Eq. 8 of \citet{Yuan2014}, $v_r=-1.1\times10^{10}\alpha\left(r/R_\mathrm{S}\right)^{-1/2}$ cm s$^{-1}$; (8) $c^2$-scaled Bernoulli parameter at $R_\mathrm{c}$ where the Bernoulli parameter is defined as Eq. \ref{eq:be}.
\end{table*}

\section{Results} \label{sec:res}

To characterize the accretion flow, we define the inflow, outflow, net mass accretion rates and the advected fraction at the inner boundary as following integrals,
\begin{align}
    \dot{M}_\mathrm{in}(r,t)&=-2\pi r^2\int_0^\pi{\rho\min{\left(v_r,0\right)}\sin\theta d\theta},\label{eq:in}\\
    \dot{M}_\mathrm{out}(r,t)&=2\pi r^2\int_0^\pi{\rho\max{\left(v_r,0\right)}\sin\theta d\theta},\label{eq:out}\\
    \dot{M}_\mathrm{acc}(r,t)&=\dot{M}_\mathrm{in}(r,t)-\dot{M}_\mathrm{out}(r,t),\label{eq:net}\\
    f_\mathrm{acc}(t)&=\dot{M}_\mathrm{acc}(r_\mathrm{in},t)/\dot{M}_\mathrm{fb}(t).\label{eq:frac}
\end{align}
It should be noted that only the gas with $v_r>0$ and $Be>0$ can arrive at infinity to form the 'true' outflows, while the gas with $Be<0$ will eventually be re-accreted. In this work, the term 'wind' specifically points to the 'true' outflows and the term 'outflow' defined as Eq. 7 includes both the wind and the gas with $v_r>0$ and $Be<0$. The Bernoulli parameter is defined as follows,
\begin{equation}
    Be=\frac{v^2}{2} + \frac{\gamma p}{\left(\gamma-1\right)\rho} - \frac{GM}{r-R_\mathrm{S}}.\label{eq:be}
\end{equation}

\subsection{General results} \label{subsec:general}

\begin{figure*}
\includegraphics[width=\textwidth]{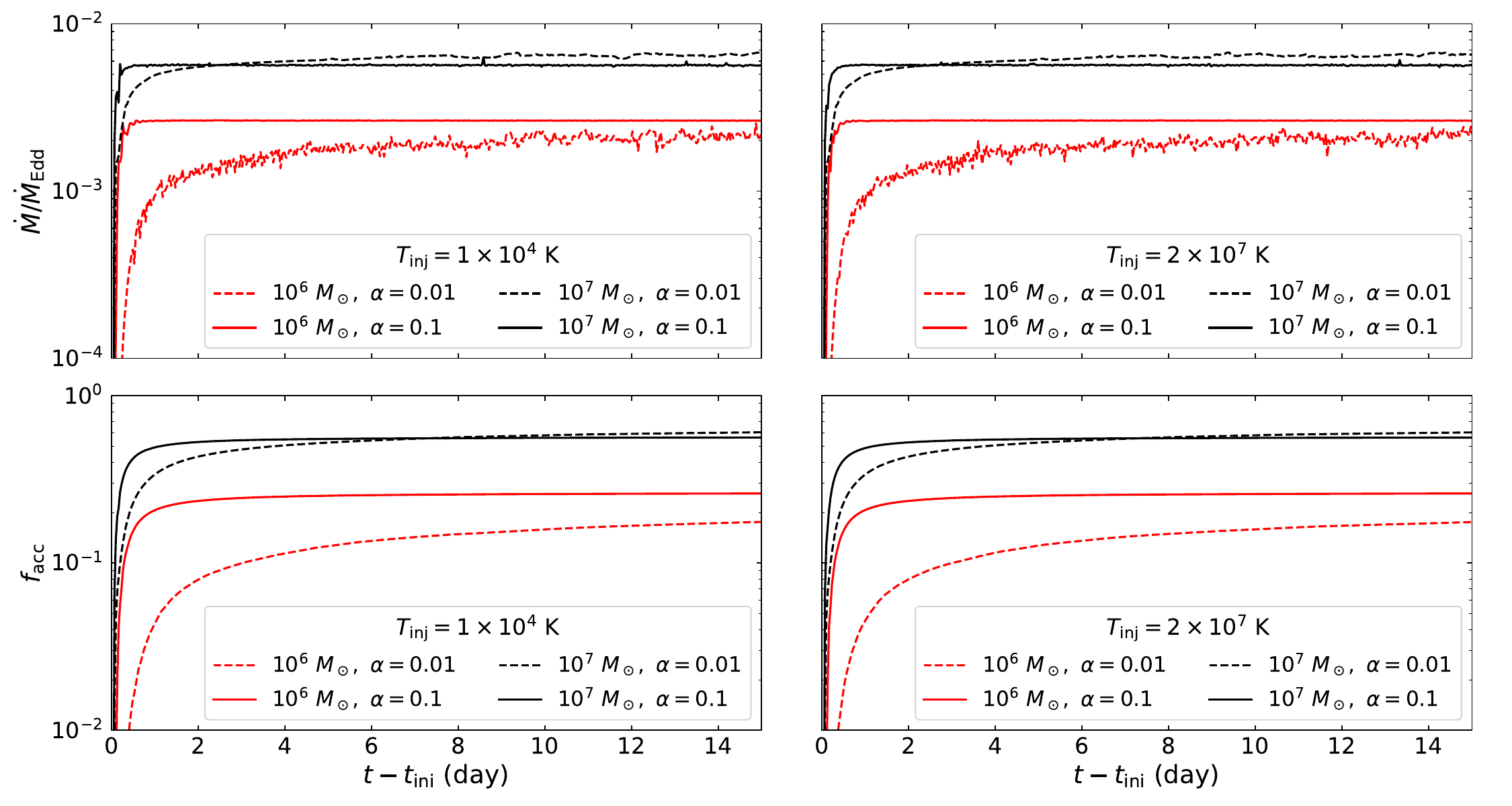}
\caption{Accretion rates (Eq. \ref{eq:net}, upper panels) and the accreted fractions of the injected mass (Eq. \ref{eq:frac}, lower panels) at the inner boundary for models with $M=10^6\ M_\odot$ (red) and $10^7\ M_\odot$ (black), $\alpha=0.01$ (dashed) and 0.1 (solid), $T_\mathrm{inj}=1\times10^4$ K (left panel) and $2\times10^7$ K (right panel).
\label{fig:acc_t}}
\end{figure*}

\begin{figure*}
\includegraphics[width=\textwidth]{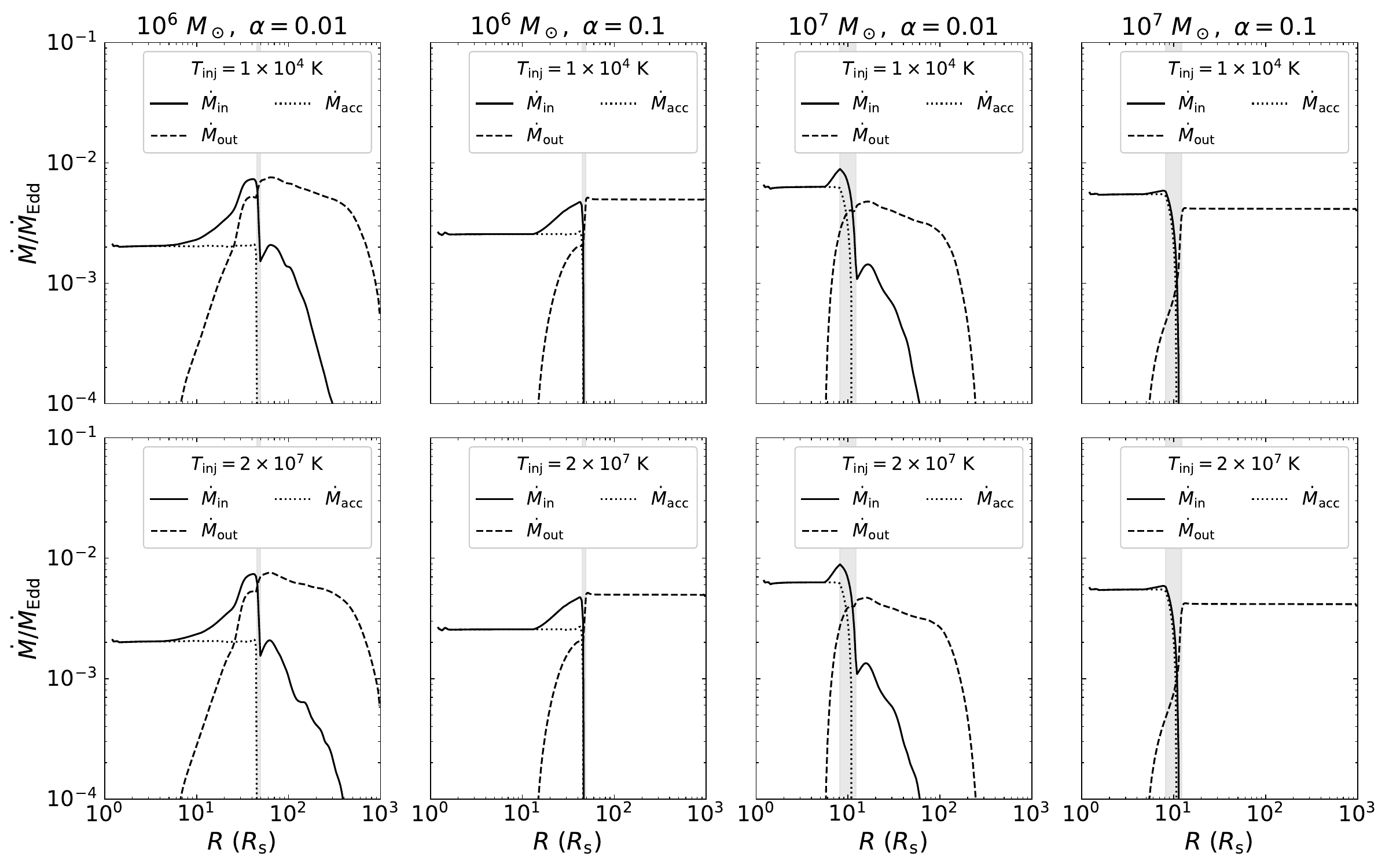}
\caption{Time-averaged radial profile of inflows and outflows from 10 to 15 days after $t_\mathrm{ini}$. The inflow, outflow and net accretion rates, derived using Eq. \ref{eq:in}, Eq. \ref{eq:out} and Eq. \ref{eq:net}, are represented by solid, dashed, and dotted lines, respectively. The gas injection zones ($R_\mathrm{c}-2R_\mathrm{S}\leq r\leq R_\mathrm{c}+2R_\mathrm{S}$) are marked by gray shadows.
\label{fig:acc_r}}
\end{figure*}

In the sub-Eddington regime (e.g., $0.01\dot{M}_\mathrm{Edd}$), the fallback rate takes decades to decline by an order of magnitude. In contrast, the viscous timescales $t_\mathrm{acc}$ at $R_\mathrm{c}$ are only a few days, as presented in Table \ref{tab:para}. Although the debris is injected into the initial tenuous background gas, the accretion flows can rapidly evolve to the quasi-steady stable state on the viscous timescale, as shown in Figure \ref{fig:acc_t}. Therefore, the quasi-steady approximation is effective for modeling TDEs in the sub-Eddington regime. In all cases, the accreted fraction remains significantly below unity, indicating the presence of outflows. Furthermore, we computed the time-averaged radial profiles of inflow and outflow, as Figure \ref{fig:acc_r} shows. It demonstrates that outflows exist ubiquitously, especially outside injection regions, and are highly influenced by the black hole mass and viscosity parameter.

More massive black holes exhibit systematically higher accreted fractions, primarily because debris is injected at a closer position scaled by $R_\mathrm{S}$ making gas more tightly gravitationally bound. Consequently, with a similar efficiency of angular momentum transfer, a smaller fraction of the debris can escape from the gravitational potential of a more massive black hole to form outflows. 

With a higher viscosity parameter, inflows transfer angular momentum outward more efficiently, thus launching stronger outflows. For $\alpha = 0.1$, the outflow profile is nearly flat outside $R_\mathrm{c}$, while almost all the inflow is confined inside the injection radius. The outflow rates remain constant outside the injection region, indicating that these outflows are unbound and will not fall back leading to ultra-low inflow rates outside $R_\mathrm{c}$. In contrast, for $\alpha = 0.01$, the inflow extends beyond $R_\mathrm{c}$ and the outflow declines at several hundred $R_\mathrm{S}$. The outflow rates at the outer boundary are much lower than the inflow rates at the inner boundary. These features imply that most of outflows are gravitationally bound when angular momentum transfer is relatively inefficient.

The temperature of the injected debris has only a weak influence on the structure of the accretion flow, as Figure \ref{fig:acc_t} and Figure \ref{fig:acc_r} show. This is because the gravitational energy released during accretion dominates over the initial internal energy of the gas. Consequently, we restrict our subsequent analysis to the cases with $T_\mathrm{inj} = 1\times10^4\ \mathrm{K}$.

\subsection{Structure of wind} \label{subsec:wind}

\begin{figure*}
\includegraphics[width=\textwidth]{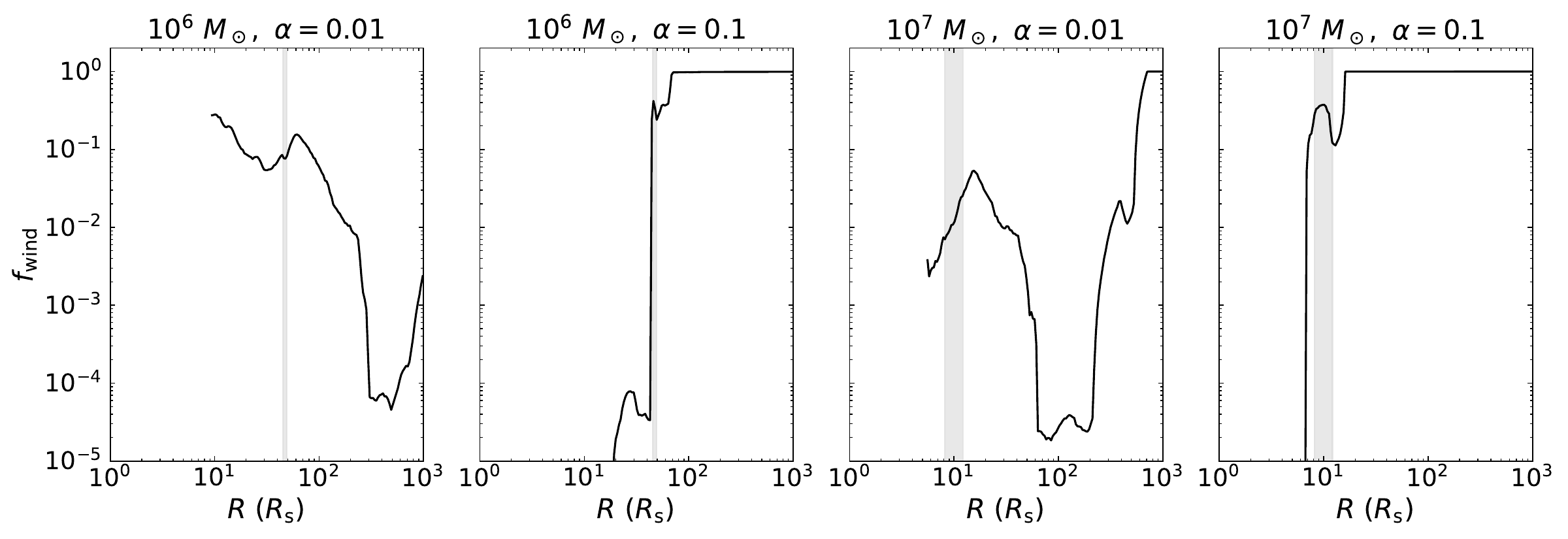}
\caption{Time-averaged radial profile of wind fraction in outflows (see Eq. \ref{eq:frac_wind}) from 10 to 15 days after $t_\mathrm{ini}$. The gas injection zones ($R_\mathrm{c}-2R_\mathrm{S}\leq r\leq R_\mathrm{c}+2R_\mathrm{S}$) are marked by gray shadows.
\label{fig:wind_frac}}
\end{figure*}

Our results demonstrate that outflows are ubiquitous in the highly sub-Eddington phase of TDEs. However, only the fraction of the outflow with a positive Bernoulli parameter ($Be>0$) can escape the black hole potential to form unbound winds. The mass flux, radial momentum flux, kinetic energy flux and the fraction of wind are therefore defined as,
\begin{align}
    \dot{M}_\mathrm{wind}&=2\pi r^2\int_0^\pi{\max\left(\frac{Be}{|Be|},0\right)\rho\max{\left(v_r,0\right)}\sin\theta d\theta},\\
    \dot{p}_\mathrm{r,wind}&=2\pi r^2\int_0^\pi{\max\left(\frac{Be}{|Be|},0\right)\rho v_r\max{\left(v_r,0\right)}\sin\theta d\theta},\\
    \dot{E}_\mathrm{wind}&=2\pi r^2\int_0^\pi{\max\left(\frac{Be}{|Be|},0\right)\rho\frac{v_r^2}{2}\max{\left(v_r,0\right)}\sin\theta d\theta},\\
    f_\mathrm{wind}&=\dot{M}_\mathrm{wind}/\dot{M}_\mathrm{out}.\label{eq:frac_wind}
\end{align}
Figure \ref{fig:wind_frac} presents the wind fraction in the outflows, representing the unbound component. For $\alpha = 0.1$, nearly all outflows are unbound. In contrast, for $\alpha = 0.01$, most of the outflow exhibits a negative Bernoulli parameter and cannot form the true winds. Due to these fundamental differences, the subsequent analysis of the wind structure focuses on models with $\alpha = 0.1$.

\begin{figure*}
\includegraphics[width=\textwidth]{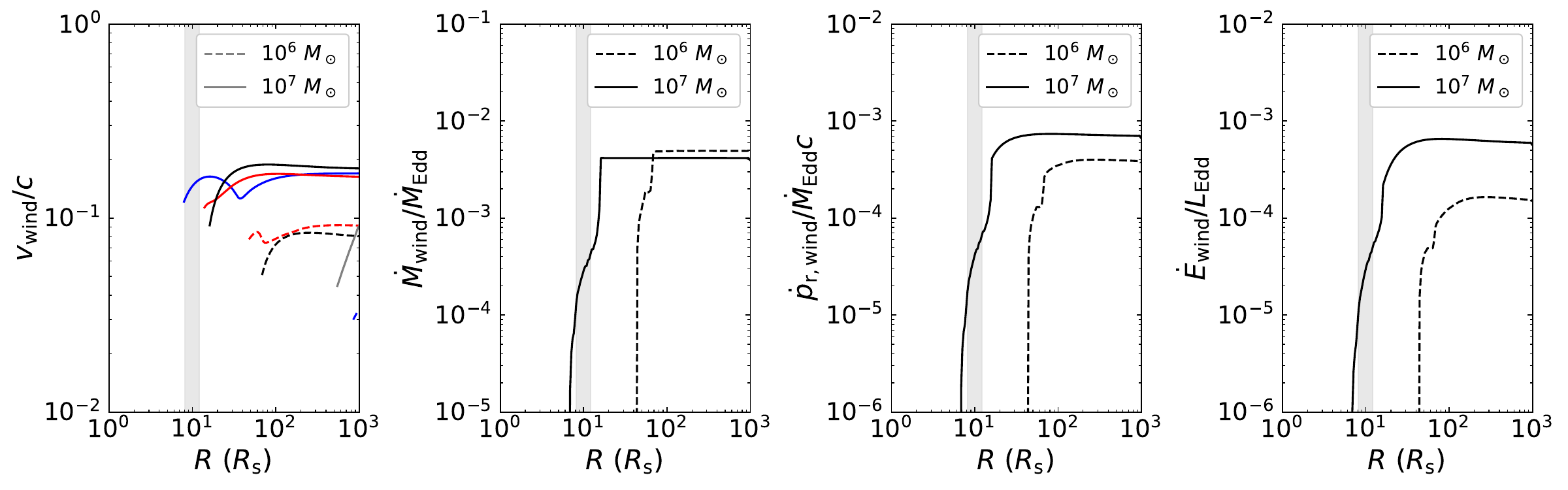}
\caption{Time-averaged radial profiles of radial velocity, mass flux, radial momentum flux and kinetic energy flux of winds for models with $M=10^6\ M_\odot$ (dashed lines) and $10^7\ M_\odot$ (solid lines), $\alpha=0.1$ and $T_\mathrm{inj}=1\times10^4$ K from 10 to 15 days after $t_\mathrm{ini}$. Profiles of radial velocity at $0^\circ$, $30^\circ$, $60^\circ$ and $90^\circ$ are marked by gray, blue, red and black lines, respectively. The gas injection zones ($R_\mathrm{c}-2R_\mathrm{S}\leq r\leq R_\mathrm{c}+2R_\mathrm{S}$) are marked by gray shadows. The radial velocity, mass flux, radial momentum flux and kinetic energy flux are scaled by $c$, $\dot{M}_\mathrm{Edd}$, $\dot{M}_\mathrm{Edd}c$ and the Eddington luminosity $L_\mathrm{Edd}$, respectively. 
\label{fig:wind_r}}
\end{figure*}

Figure \ref{fig:wind_r} shows the time-averaged radial profiles of radial velocity, as well as mass, momentum, and kinetic energy fluxes for the winds in the $\alpha=0.1$ models. Winds with velocities of approximately $0.1c$ develop away from the polar axis for both $10^6\ M_\odot$ and $10^7\ M_\odot$ black holes. Because the debris is injected from a closer region for more massive black holes, a larger fraction of the debris is accreted and the unbound gas must reach a higher velocity to escape the deeper gravitational potential. As a result, the wind from the $10^6\ M_\odot$ black hole exhibits a higher mass flux, but its lower velocity reduces the momentum and kinetic energy fluxes. Overall, hot accretion flows around high mass black holes can launch more relativistic, powerful but tenuous winds in TDEs. 

\begin{figure*}
\includegraphics[width=\textwidth]{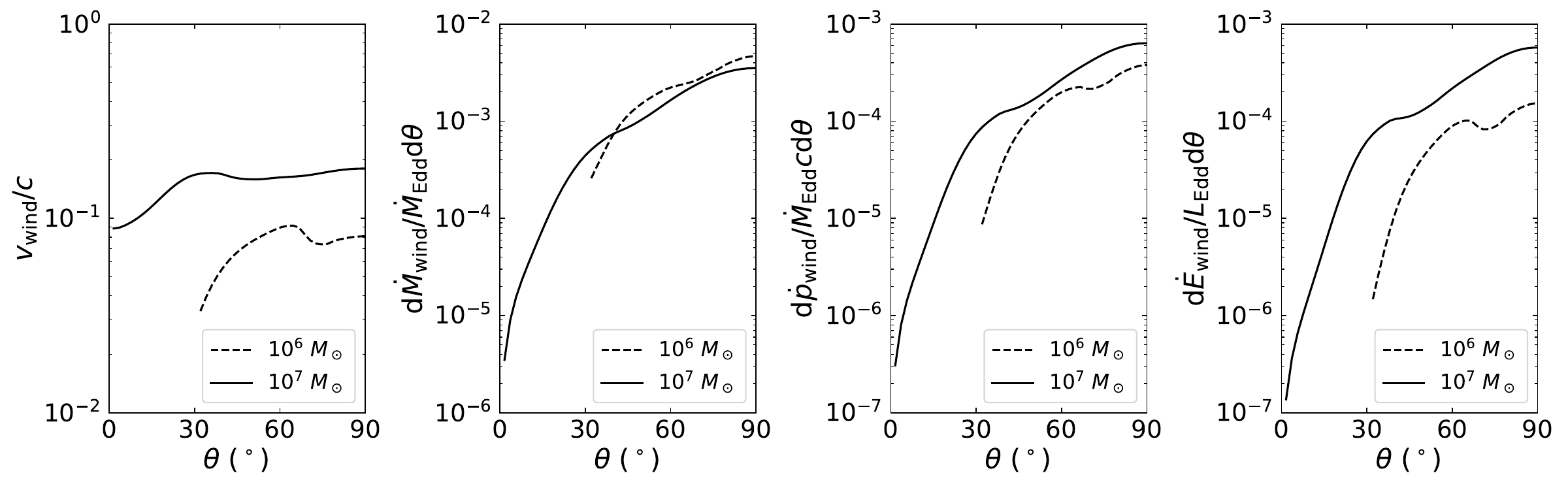}
\caption{Time-averaged angular distributions at 1000 $R_\mathrm{S}$ of the radial velocity, mass flux, radial momentum flux and kinetic energy flux of winds for models with $M=10^6\ M_\odot$ (dashed lines) and $10^7\ M_\odot$ (solid lines), $\alpha=0.1$ and $T_\mathrm{inj}=1\times10^4$ K from 10 to 15 days after $t_\mathrm{ini}$. Each quantity is scaled identically to those in Figure \ref{fig:wind_r}.
\label{fig:wind_angle}}
\end{figure*}

Figure \ref{fig:wind_angle} presents the time-averaged angular profiles of the velocity and fluxes of the winds with $\alpha=0.1$. For both $10^6\ M_\odot$ and $10^7\ M_\odot$ black holes, unbound winds are launched over a wide angular range (exceeding $40^\circ$ from the pole for a $10^6\ M_\odot$ black hole), extending to lower latitudes for more massive black holes at larger radii. In both cases, the wind carries more mass, momentum, and kinetic energy at more edge-on directions.

Furthermore, the structure of the wind launched in the sub-Eddington regime differs markedly from that in the super-Eddington phase of TDEs. \citet{Bu2023b} performed hydrodynamical simulations to investigate super-Eddington TDE winds for $10^6$ and $10^7\ M_\odot$ black holes with $\alpha=0.1$. Those super-Eddington winds reach velocities of $0.1-0.7c$, significantly faster than the sub-Eddington winds found here for the same $\alpha$. Fueled by a much extensive gas supply, the super-Eddington winds carry a kinetic energy of $(2-6.5)\times10^{44}$ erg s$^{-1}$, whereas the sub-Eddington winds attain only a few $\times 10^{-4}\ L_\mathrm{Edd}$, as shown in Figure \ref{fig:wind_r}. The sub-Eddington wind is primarily launched at edge-on angles, while the super-Eddington wind is more powerful at intermediate angles.

\subsection{Convective bound outflow}

\begin{figure*}
\includegraphics[width=\textwidth]{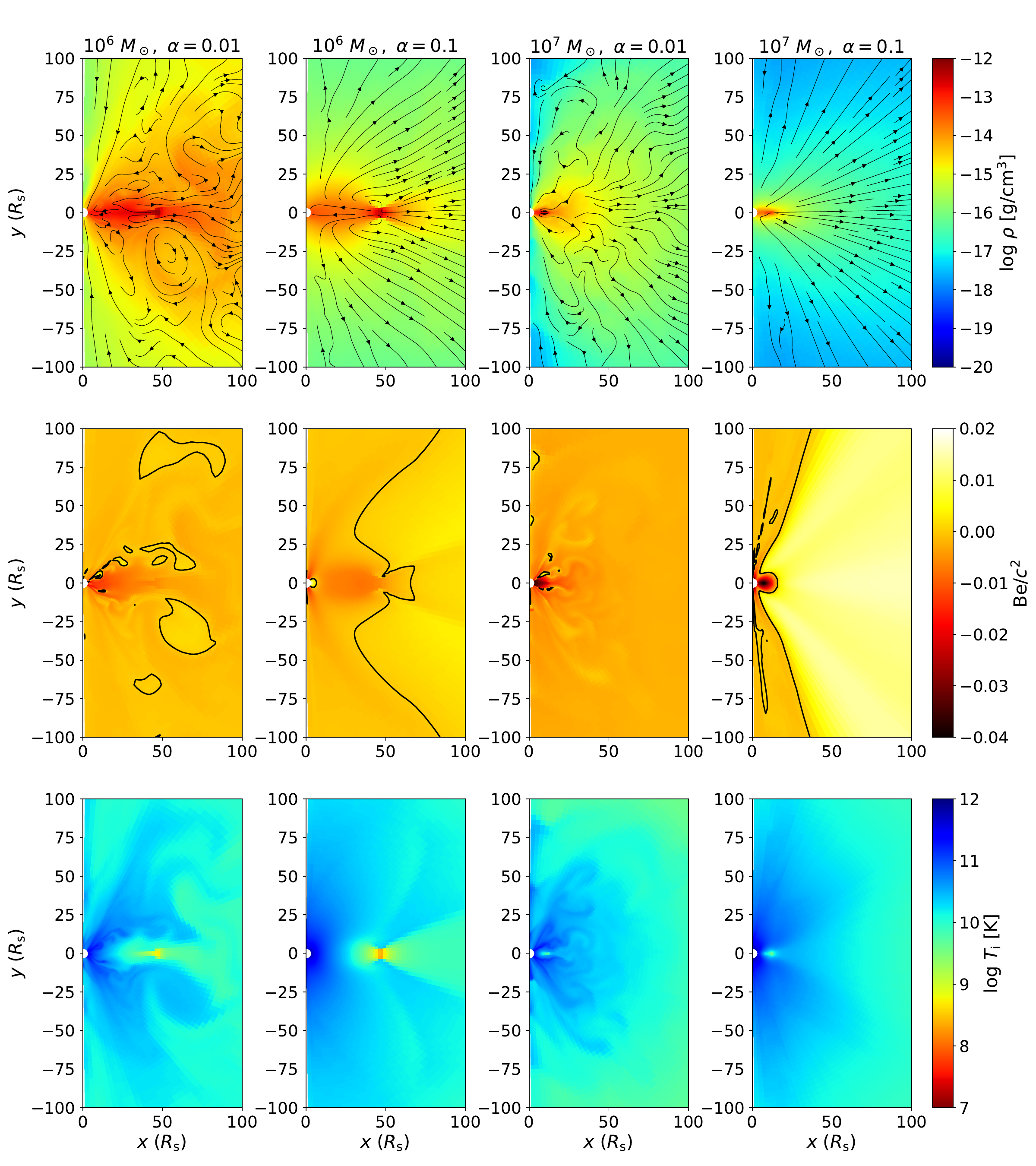}
\caption{Distributions of density (upper panel), Bernoulli number (middle panel) and temperature (lower panel) within 100 $R_\mathrm{S}$ for models with $M=10^6\ M_\odot$ and $10^7\ M_\odot$, $\alpha=0.01$ and 0.1, and $T_\mathrm{inj}=1\times10^4$ K at 15 days after $t_\mathrm{ini}$. Streamlines are indicated by black soild arrows in the upper panel. The $Be=0$ contours are marked by black solid lines in the middle panel.
\label{fig:be}}
\end{figure*}

\begin{figure*}
\includegraphics[width=\textwidth]{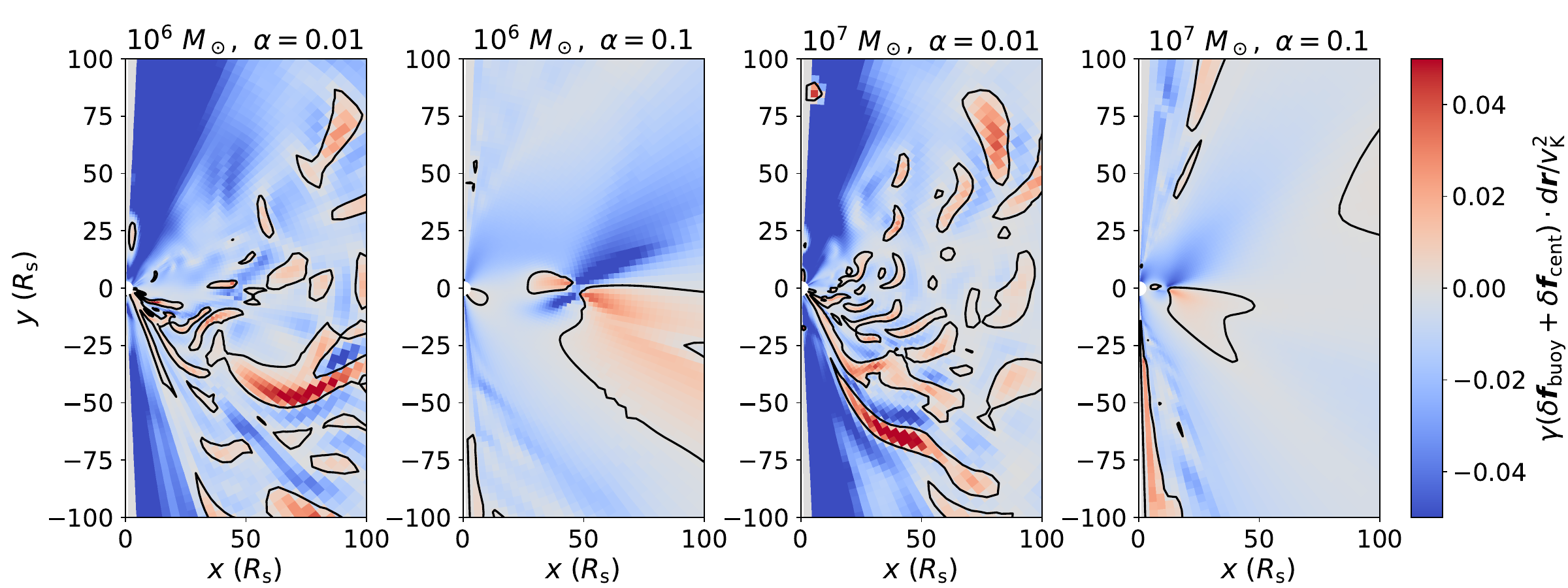}
\caption{Distributions of the $v_\mathrm{K}^2$-normalized Høiland criterion within 100 $R_\mathrm{S}$ for models with $M=10^6\ M_\odot$ and $10^7\ M_\odot$, $\alpha=0.01$ and 0.1, and $T_\mathrm{inj}=1\times10^4$ K at 15 days after $t_\mathrm{ini}$. The $(\delta\bm{f}_\mathrm{buoy}+\delta\bm{f}_\mathrm{cent})\cdot d\bm{r}=0$ contours are marked by black solid lines.
\label{fig:cdaf}}
\end{figure*}

As the results above demonstrate, outflows in the sub-Eddington phase of TDEs exhibit distinct properties depending on the viscosity parameter. In particular, the radial and angular distributions of velocity and mass flux strongly support the presence of convection in the cases with $\alpha=0.01$. To better understand the mechanisms governing outflows in the late stages of TDEs, we visualize the density, streamlines, Bernoulli parameter, temperature and the Høiland criterion \citep[e.g.,][]{Tassoul1978,Begelman1982} in the $r–\theta$ plane at 15 days after $t_\mathrm{ini}$, as shown in Figure \ref{fig:be} and Figure \ref{fig:cdaf}. The criterion for convective stability $\gamma(\delta\bm{f}_\mathrm{buoy}+\delta\bm{f}_\mathrm{cent})\cdot d\bm{r}$ is
\begin{equation}
    \left(\nabla s\cdot d\bm{r}\right)\left(\bm{g}_\mathrm{eff}\cdot d\bm{r}\right)-\frac{2\gamma v_\phi^2}{R}\left[\nabla\left(v_\phi R\right)\cdot d\bm{r}\right]dR<0,
    \label{eq:stable}
\end{equation}
where dimensionless specific entropy $s=\ln{\left(p/\rho^\gamma\right)}$, effective gravity $\bm{g}_\mathrm{eff}=-\hat{r}v_\mathrm{K}^2/r+\hat{R}v_\phi^2/R$, Keplerian velocity $v_\mathrm{K}=\sqrt{GM/(r-R_\mathrm{S})}$, cylindrical radius $R=r\sin\theta$, displacement vector $d\bm{r}=dr\hat{r}+rd\theta\hat{\theta}$, and unit direction vectors $\hat{r}$, $\hat{\theta}$ and $\hat{R}$ in $r$, $\theta$ and $R$ direction, respectively. The gas in models with $\alpha=0.01$ is markedly more turbulent than that with $\alpha=0.1$. When $\alpha=0.01$, clear vortex structures are present around both $10^6$ and $10^7\ M_\odot$ black holes, while most of these eddies are gravitational bound with negative Bernoulli parameters. The vortical patterns are also presented in the temperature profiles. And some of these regions exhibit convective instability as Figure \ref{fig:cdaf} shows. These support the presence of buoyancy-driven convection in the cases of $\alpha=0.01$. When $\alpha=0.1$, the true winds are launched at angles $\theta > 40^\circ$ with comparatively lower temperatures compared with inflows. However, a region near the outside of the injection zone with a positive Høiland criterion extends dozens of $R_\mathrm{S}$ in each case of $\alpha=0.1$, while only a few small convective eddies exist near the injection region as the upper panel of Figure \ref{fig:be} shows. This is because the Eq.(\ref{eq:stable}) is derived from the inviscid fluid approximation \citep{Begelman1982}. For the high viscous cases with $\alpha=0.1$, most of vortex structures have been smoothed, indicating the overestimation of convective instability from Eq.(\ref{eq:stable}).

\begin{figure*}
\includegraphics[width=\textwidth]{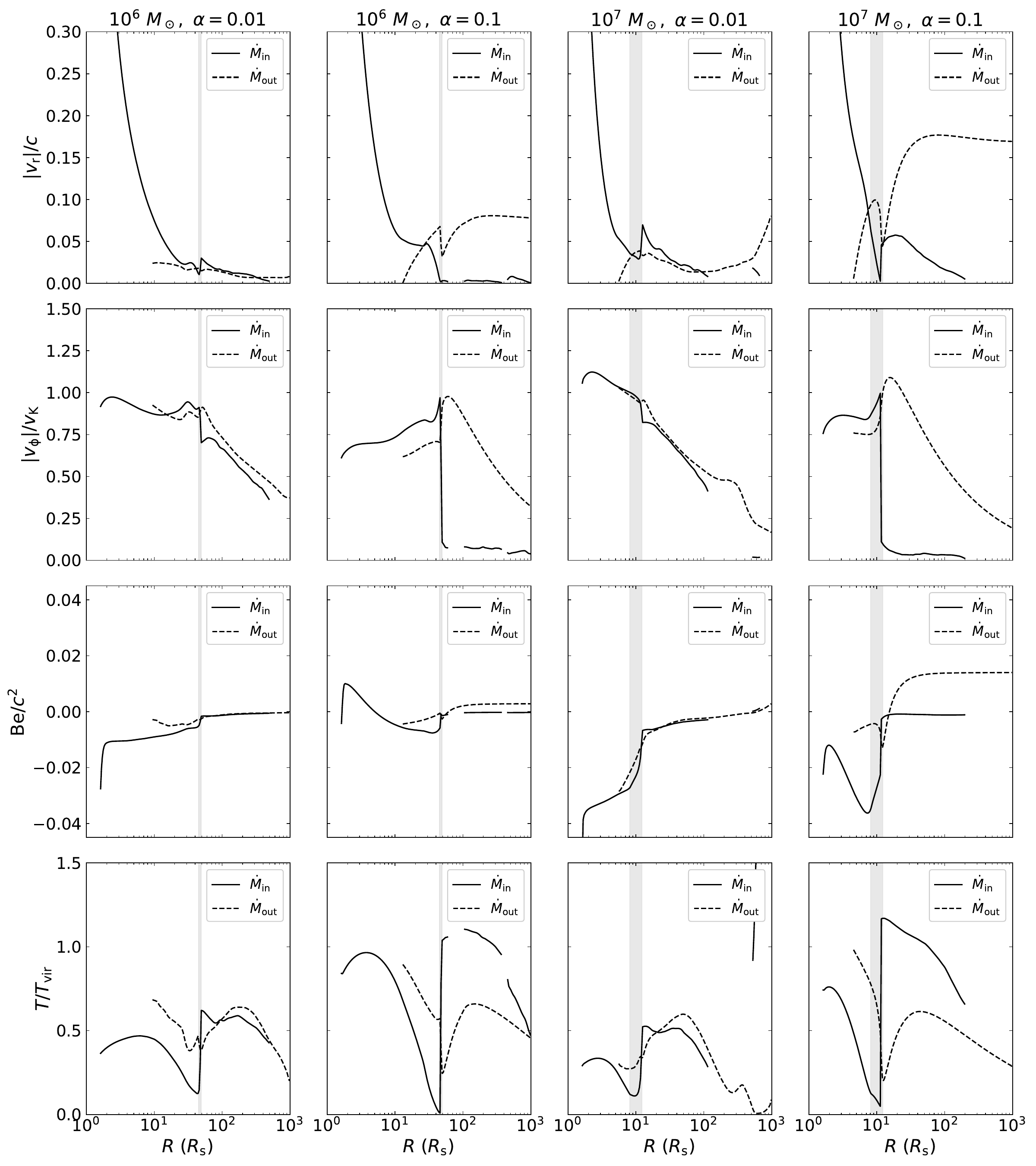}
\caption{Time-averaged radial profiles from 10 to 15 days after $t_\mathrm{ini}$ of the mass-flux-weighted radial velocity, rotation velocity, Bernoulli parameter and temperature for the inflow (solid) and outflow (dashed) for models with $M=10^6\ M_\odot$ and $10^7\ M_\odot$, $\alpha=0.01$ and 0.1, and $T_\mathrm{inj}=1\times10^4$. Radial velocity, rotation velocity, Bernoulli number and temperature are scaled by $c$, $v_\mathrm{K}$, $c^2$ and the virial temperature $T_\mathrm{vir}$, respectively. The gas injection zones ($R_\mathrm{c}-2R_\mathrm{S}\leq r\leq R_\mathrm{c}+2R_\mathrm{S}$) are marked by gray shadows.
\label{fig:outflow}}
\end{figure*}

Finally, we calculate the mass-flux-weighted radial velocity, Bernoulli number and temperature. For inflow and outflow, the mass-flux-weighted quantity $q$ is defined as follows,
\begin{align}
    <q(r,t)>_\mathrm{in}&=\frac{2\pi r^2\int_0^\pi{\rho q\min{\left(v_r,0\right)}\sin\theta d\theta}}{2\pi r^2\int_0^\pi{\rho\min{\left(v_r,0\right)}\sin\theta d\theta}}, \\
    <q(r,t)>_\mathrm{out}&=\frac{2\pi r^2\int_0^\pi{\rho q\max{\left(v_r,0\right)}\sin\theta d\theta}}{2\pi r^2\int_0^\pi{\rho\max{\left(v_r,0\right)}\sin\theta d\theta}}.
\end{align}
The time-average treatment between 10 and 15 days after $t_\mathrm{ini}$ is applied to reduce the influence of fluctuations. As shown in Figure \ref{fig:outflow}, the inflow and outflow with $\alpha=0.1$ exhibit clearly distinct profiles in the radial velocity, rotation velocity, Bernoulli number and temperature, which identify the outflows as true unbound winds. It is consistent with the typical AGN and X-ray binary (XRB) scenarios \citep{Yuan2012b} but differs in the geometrical configuration. The inflow and outflow with $\alpha=0.01$ exhibit similar velocity and Bernoulli parameter over radial distances from several tens to several hundred $R_\mathrm{S}$. These also show that the inflows and outflows belong to the inward and outward portions of convective eddies, respectively.

\begin{figure}
\includegraphics[width=\columnwidth]{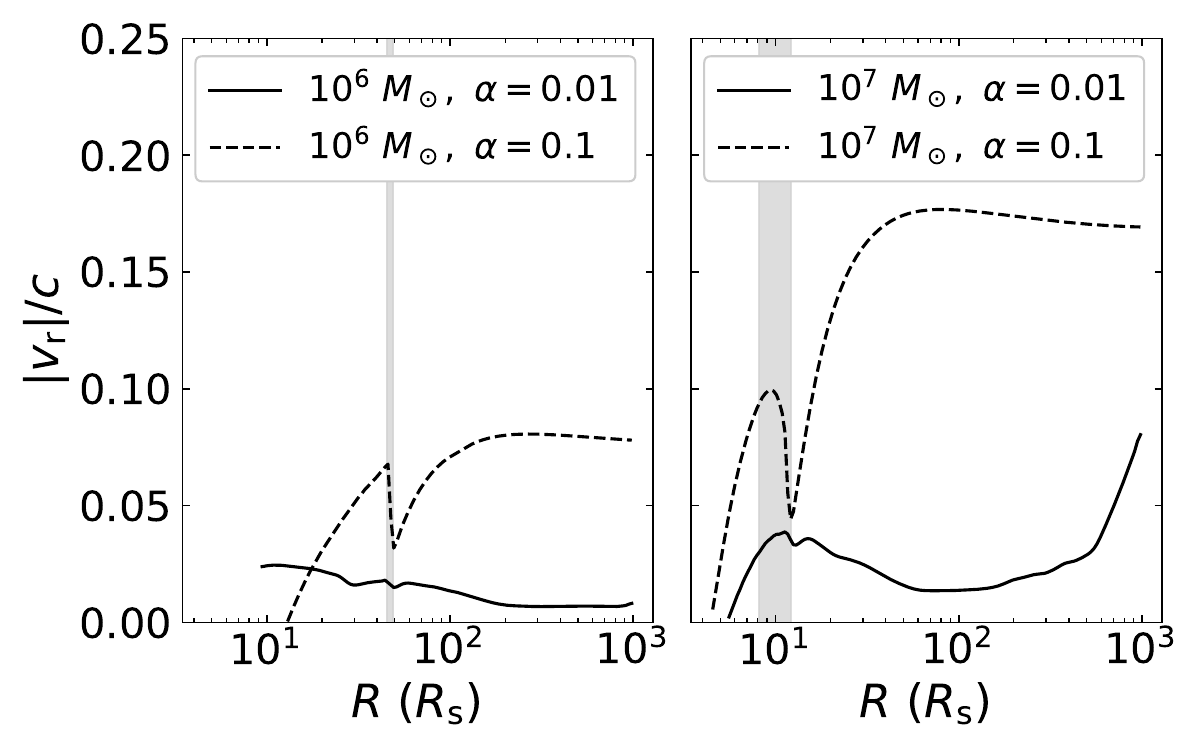}
\caption{Time-averaged radial profiles from 10 to 15 days after $t_\mathrm{ini}$ of the mass-flux-weighted $c$-scaled radial velocity of the outflow for models with $M=10^6\ M_\odot$ (left panel) and $10^7\ M_\odot$ (right panel), $\alpha=0.01$ (solid) and 0.1 (dashed), and $T_\mathrm{inj}=1\times10^4$. The gas injection zones ($R_\mathrm{c}-2R_\mathrm{S}\leq r\leq R_\mathrm{c}+2R_\mathrm{S}$) are marked by gray shadows.
\label{fig:compare}}
\end{figure}

A higher viscosity parameter facilitates more efficient angular momentum transfer within the accretion flow. Consequently, a larger fraction of gas can acquire sufficient angular momentum to escape the black hole gravitational potential, leading to a pronounced differentiation between inflow and outflow. This can be seen from the direct comparison of the radial velocity in outflows with different viscosity parameters, as shown in Figure \ref{fig:compare}. The outflows with a higher viscosity parameter can move outward much faster. For a more massive black hole, the gas is injected at a smaller circularization radius. Therefore, winds need to have a higher velocity to escape from the gravitational potential well. On the other hand, with a large viscosity parameter, accretion flows become laminar, suppressing convection \citep{Igumenshchev1999,Igumenshchev2000}. However, \citet{Yuan2012b} investigated three hydrodynamical models with $\alpha=0.01$ and found that all exhibited ADIOS-like features, i.e., the presence of unbound winds above the accretion flows. These models employed different initial conditions, i.e. a rotating torus \citep[Model A][]{Stone1999}, a global hot accretion flow solution (Model B), and gas injected at an outer boundary (Model C). Models A and B initially had negative Bernoulli parameters. In contrast, the stellar debris in a TDE is injected from an extremely close pericenter distance, placing mass deep into the gravitational potential well. This results in the fact that the variations in the inflow and outflow rates at $\alpha=0.01$ in our TDE models are caused by some convective bound outflows, rather than by the conventional CDAF scenario. Furthermore, Models A and B in \citet{Yuan2012b} were in hydrostatic equilibrium and largely supported by gas pressure, while the stellar debris in a TDE is initially too cold to provide significant pressure support. Even after a hot accretion flow forms, the temperatures in our models B6$\alpha$001T4 and B7$\alpha$001T4 remain lower than in Model A of \citet{Yuan2012b}. 

\section{Observational applications}\label{sec:obs}

We found that the winds from hot accretion flows in TDEs depend on the viscosity parameter $\alpha$. Although the complexity in turbulence and magnetic field makes the origin of viscosity in accretion flow still an open question, many works have made efforts to constrain the viscosity parameter in different ways. The magnetohydrodynamic simulations give a wide range of $\alpha$ from $10^{-3}$ to 1 \citep[e.g.,][]{Stone1996,Machida2000,Hirose2006,Hawley2011,McKinney2012,Bai2013}. By estimating the timescale of the observed variability, \citet{Starling2004} derived the lower limit of $\alpha$ at 0.01 in AGNs with $0.1\leq L/L_\mathrm{Edd}\leq1$ and \citet{King2007} suggested $\alpha$ in the range of 0.1 to 0.4. By the spectral comparison between model and observation, many works obtained $\alpha\geq0.1$ in AGNs \citep[e.g.,][]{Qiao2018} and XRBs \citep[e.g.,][]{Narayan1996,Xie2016,WangYL2025}, and our previous work constrained the $\alpha\sim0.03$ in AGNs with $L>0.1L_\mathrm{Edd}$ \citep{Liu2025b}. So, we focused on the cases with $\alpha=0.01$ and 0.1 in this work. However, understanding the viscosity in TDEs would be more difficult due to the interaction between the magnetic field in the disrupted star and the initial field near the black hole. In this work, we provide a new way to constrain the viscosity parameter through the winds in the late-stage of TDEs.

Disk-driven winds can interact with the circumnuclear medium \citep[e.g.,][]{BarniolDuran2013,Matsumoto2021,Matsumoto2024,Cendes2024,Zhou2024,Mou2025b} or with dense clouds \citep[e.g.,][]{Mou2022,Bu2023a,Lei2024,Zhuang2025,Yang2025,Mou2025a} to produce radio emission. Late radio emission has been observed in many TDEs \citep[e.g.,][]{Alexander2020,Horesh2021,Cendes2022,Cendes2024,Goodwin2022,Perlman2022,Sfaradi2022,Wang2023,Wang2025,Anumarlapudi2024,Zhang2024b,Lin2025}, indicating the ubiquitous winds in TDEs. Recently, \citet{Alexander2025} reported delayed radio brightening in 12 TDEs and a second radio brightening of AT2019dsg at around 1000 days, consistent with an accretion rate below $0.03\dot{M}_\mathrm{Edd}$ as estimated from Modular Open Source Fitter for Transients \citep[MOSFiT,][]{Guillochon2018,Mockler2019} modeling. 

Among them, ASASSN-14ae, AT2018fyk and AT2019teq, align with the Fundamental Plane of black hole activity at late times. Therefore, their ultra-late radio brightening is likely produced by sub-Eddington winds from hot accretion flows. Their radio and X-ray luminosities around $10^3$ days after discovery are approximately $10^{37}$–$10^{38}\ \mathrm{erg\ s^{-1}}$ and $\sim10^{41}\ \mathrm{erg\ s^{-1}}$, respectively \citep{Alexander2025}. Using the $M$–$\sigma$ relation, the black hole masses for ASASSN-14ae and AT2019teq are estimated as $10^{6.0\pm0.3}\ M_\odot$ and $10^{6.3\pm0.3}\ M_\odot$, respectively, while AT2018fyk appears to host a more massive black hole of $10^{8.0\pm0.3}\ M_\odot$. However, tidal disruption of a solar-type star by a black hole of this mass is unlikely unless it possesses high spin. X-ray spectra analysis indicates that AT2018fyk is associated with a super massive black hole binary with a secondary black hole of $2.7^{+0.5}_{-1.5}\times10^5M_\odot$ \citep{Wen2024}. As shown in Figure \ref{fig:wind_r}, the sub-Eddington winds in SMBH-TDEs can carry kinetic energy up to a few $10^{-4}L_\mathrm{Edd}$ with $\alpha=0.1$, which is sufficient to power $10^{37}$–$10^{38}\ \mathrm{erg\ s^{-1}}$ radio emission. These radio emissions from sub-Eddington winds implies that the accretion flows in ASASSN-14ae, AT2019teq and AT2018fyk likely have a viscosity parameter of $0.01<\alpha<0.1$. In addition, the hot accretion flow itself produces broadband emission mainly from the infrared to hard X-ray bands with an efficiency of $\sim0.01$ for $\dot{M}=0.01\dot{M}_\mathrm{Edd}$ \citep[e.g.,][]{Narayan1995b}, which is comparable to the X-ray luminosity of ASASSN-14ae, AT2018fyk and AT2019teq. Hence, their X-ray band is likely dominated by emission from the hot accretion flow. 

TDEs on intermediate-mass black holes (IMBHs) evolve more slowly than those on supermassive black holes. As Figure \ref{fig:fb} indicates, the typical TDE rate of $\sim 10^{-4}$ yr$^{-1}$ per galaxy implies that IMBHs in the range of $10^4$–$10^5\ M_\odot$ could sustain persistent, low-level activity induced by TDEs. Because a main-sequence star is disrupted at a much larger distance from an IMBH, the accreted fraction of the hot accretion flow is expected to be extremely low (see Figure \ref{fig:acc_t}) and winds can more readily form. The long-lived sub-Eddington winds from IMBH-TDEs show that the TDE-driven feedback should play a role in dwarf galaxies as well as the galaxy formation in early universe. Therefore, radio and X-ray surveys of globular clusters and dwarf galaxies offer a promising way to discover more IMBHs, which is also supported by the preliminary analysis for the \emph{Square Kilometre Array} \citep{Liodakis2022} and current observations by the \emph{Einstein Probe} \citep{Yuan2022,Yuan2025,Zhang2025,Jin2025,Shu2025,Li2025}.

\section{Conclusion} \label{sec:con}

In this work, we investigate winds launched from hot accretion flows during the sub-Eddington phase of TDEs using hydrodynamical simulations. To explore the effects of black hole mass, angular momentum transfer efficiency and stellar debris temperature, we employ eight distinct models with black hole masses of $M = 10^6$ and $10^7 M_\odot$, viscosity parameters $\alpha = 0.01$ and $0.1$, and injection temperatures $T_\mathrm{inj} = 1\times10^4$ K and $2\times10^7$ K. All models correspond to a solar-type star disrupted on a $\beta = 1$ orbit, starting at an accretion rate of $0.01 \dot{M}_\mathrm{Edd}$. We find that the accretion flows reach a steady state within approximately one viscous timescale, validating the quasi-steady approximation for TDEs in the sub-Eddington regime.

Outflows are highly influenced by the black hole mass and viscosity parameter. For more massive black holes, debris is injected at a smaller scaled radius $R_\mathrm{c}/R_\mathrm{S}$, resulting in a lower Bernoulli parameter. Consequently, a larger fraction of the debris is accreted and any escaping wind requires a higher velocity. The temperature of the injected debris has a negligible influence on the accretion flow evolution because the gravitational energy released during accretion dominates over the initial internal energy of the gas. The efficiency of angular momentum transfer largely dictates the outflow structure. For $\alpha = 0.1$, the mildly-relativistic unbound winds ($\sim 0.1c$) are launched predominantly from the injection region along the equatorial directions, with a kinetic energy of $\sim10^{-4} L_\mathrm{Edd}$. In contrast, outflows are convectively bound for $\alpha = 0.01$, while the outflows from hot accretion flows in AGNs and XRBs are still unbound with this level of angular momentum effciency \citep[e.g.,][]{Yuan2012b}. This likely stems from the unique initial conditions of TDEs, i.e., initially cold stellar debris injected at a low Bernoulli parameter.

We also discuss potential astrophysical applications of winds from hot accretion flows in TDEs, including explanations for delayed radio brightening observed in TDEs at $\sim 10^3$ days, and the prospect of detecting IMBHs through radio and X-ray sky surveys.

\begin{acknowledgments}

We thank the anonymous referee for the helpful suggestions that improved this paper, and the authors thank Dongyue Li and Weimin Yuan for their valuable discussions on the TDE observations. We acknowledge the support by the National Natural Science Foundation of China (NSFC, grant No. 12333004) and the National Key R\&D Program of China No. 2025YFF0511100. D. Bu is supported by the National SKA Program of China (No. 2025SKA0130100) and the NSFC (grants 12192220 and 12192223). We thank the participants of the TDE FORUM (Full-process Orbital to Radiative Unified Modeling) online seminar series for their inspiring discussions.
\end{acknowledgments}

\software{ZEUS-MP \citep{Hayes2006}.}






\end{document}